\documentclass[prd,aps,preprintnumbers,nofootinbib,twocolumn]{revtex4-1}
\pdfoutput=1 

\usepackage{amsthm}
\usepackage{amssymb}
\usepackage{amsmath}
\usepackage{amscd}
\usepackage{hyperref}
\usepackage{url}
\usepackage{tabularx}
\usepackage{float}

\newtheorem{myth?}{Theorem?}

\begin{document} 
\title{Gauge anomalies of finite groups}

\author{Ben Gripaios} \email{gripaios@hep.phy.cam.ac.uk} \affiliation{Cavendish Laboratory, University of Cambridge, J.J. Thomson
Avenue, Cambridge, CB3 0HE, United Kingdom}
\begin{abstract}
We show how the theory of characters can be used to analyse an
anomaly corresponding to chiral fermions
carrying an arbitrary representation of a gauge
group that is finite, but otherwise arbitrary. By way of example, we
do this for some groups of
relevance for the study of quark and lepton masses and mixings. 
\end{abstract}
\maketitle
\onecolumngrid
\begin{center}
{\sc In Memoriam Graham Garland Ross FRS, 1944-2021}
\end{center}
\vspace{0.5cm}
\twocolumngrid
Finite symmetry groups are ubiquitous in physics, {\em e.g.} for stabilizing
particles such as the proton or dark matter, or for explaining the patterns
of masses and mixings of quarks and leptons. The lore of quantum gravity says that
they should be gauged and the lore of gauge symmetry says that they
should be free of anomalies that can arise when they act on chiral
fermions, at least if they are to be linearly realized
{\em in vacuo}.

The study of such anomalies was pioneered by Graham Ross in
collaboration with Luis Iba\~{n}ez \cite{Ibanez:1991hv}, who studied cyclic groups
by embedding them as subgroups of a spontaneously broken $U(1)$ (see also
\cite{Banks:1991xj}). The extension to an arbitrary finite abelian
group follows immediately, since such a group is isomorphic to a product of
cyclic groups, but the general case of groups that are not necessarily
abelian is still poorly understood. This is unfortunate, since groups
with irreducible representations whose dimensions exceed one, which is a {\em sine qua
  non} for applications to flavour physics, are necessarily
non-abelian. 

Here we show that one can perform a complete analysis of the
anomalies of an arbitrary finite group using the theory of characters. To
give a flavour of the power of this approach, consider the following
situation: given a group and a representation of it, there must exist
a unique
largest subgroup that is anomaly-free with respect to that
representation. By considering all representations, one generates a
list of possible anomaly-free subgroups. As the examples considered at
the end of this Letter show, this list can be found using
the theory of characters with a line or two of menial algebra. (To match
representations with subgroups in the list takes another line or two.) 

Let us begin by recalling the current state of the art, but phrasing
things in a
way which is both mathematically {\em kosher} and as general as
possible. Suppose the spacetime dimension is 4 and that the fermions
(all with the same chirality)
carry the complex
representation $\rho$ of $G$. An argument \cite{Araki:2008ek} along the lines of
Fujikawa's \cite{Fujikawa:1979ay}, shows that the transformation corresponding to
$g \in G$ is free of a mixed anomaly between $G$ and gravity iff. $\mathrm{\det} \rho (g) = \pm 1$. The
curious factor of $\pm 1$ arises ultimately from an index theorem, but
we can see that it must be present by following
  Ross and Iba\~{n}ez and considering
  $G$ to be a cyclic group arising as the linearly-realized subgroup of a spontaneously
  broken $U(1)$ that is itself anomaly-free. The freedom
  to include a minus sign then arises from the
  possibility that a single charged chiral fermion can acquire a Majorana mass
  and thus decouple at low energy. 

No further anomalies are detected by Fujikawa's argument (we will
discuss further possible anomalies at the end) unless $G$ has a non-trivial Lie algebra,
but then $G$ is no longer finite. In that case, mixed anomalies
can arise between the normal subgroup $G_0$ consisting of the
component path-connected to the
identity element in $G$ and the group of connected components $G/G_0$, which is finite
assuming $G$ is compact. A useful stepping stone in carrying over the analysis to this
more general case is to study the slightly more general condition $(\mathrm{\det} \rho (g))^m = 1$ with $m \in
\mathbb{N}$ and we shall do so in what follows. 

A recent preprint \cite{Kobayashi:2021xfs} (see also
\cite{Araki:2006sqx,Chen:2013dpa,Chen:2015aba,Talbert:2018nkq}) goes on to establish
the following: {\em (i)} the set of
$g \in G$ such that $(\mathrm{\det} \rho (g))^m = 1$ forms a
subgroup of $G$, which we call the {\em anomaly-free
  subgroup}\footnote{Italics are used to indicate that a definition is being
  given.} and
denote $G_{\rho}^m$; {\em (ii)} $G_{\rho}^m$  is normal; and {\em (iii)} $G_{\rho}^m$ contains the {\em derived
subgroup} $G'$ (being the normal subgroup generated by {\em commutators},
{\em i.e.} elements in
$\{ghg^{-1}h^{-1}|g,h \in G\}$). Moreover, the first of these three
facts implies that we may form the set of cosets $G/G_{\rho}^m$, the
second implies that $G/G_{\rho}^m$ forms a group (which we call the
{\em anomalous quotient group}), and the third implies that it is
abelian.  It is then further shown in \cite{Kobayashi:2021xfs} that the anomalous quotient
group is cyclic. 

Our point of departure is to observe that all of these facts follow
almost immediately
once one notices that the map $g \mapsto (\mathrm{\det} \rho (g))^m$
defines a homomorphism $\pi_{\rho}^m$ from $G$ to the abelian group $S^1 \in
\mathbb{C}$ of complex numbers with unit modulus, with group law given
by complex multiplication. Indeed, $G_{\rho}^m$ is then the kernel of
$\pi_{\rho}^m$, so is normal, and $G/G_{\rho}^m$ is isomorphic to the
image of $\pi_{\rho}^m$, so is isomorphic to a subgroup of $S^1$.
But all subgroups of the abelian group $S^1$ are of necessity abelian and 
all finite subgroups of $S^1$ are
moreover cyclic\footnote{Proof. Parameterize the elements of
  $S^1$ by an angle in $[0,2\pi]$ and let $\theta$ be the smallest
  element in a finite subgroup. If the subgroup is not cyclic, then it is not
  generated by $\theta$, so there must exist an element $\phi$ such
  that $n \theta < \phi < (n+1) \theta \implies 0 < \phi - n \theta <
  \theta $. So $\theta$ is not the smallest element, which is a
  contradiction.}, so $G/G_{\rho}^m$ is cyclic.

Regarding the observation in \cite{Kobayashi:2021xfs} that $G_{\rho}^m\supset G'$, this 
is implied by our observation that 
$G/G_{\rho}^m$ is abelian, but it will be useful for
what follows to
spell out the connection in more detail. To wit, we have that
$G^\prime$ is a normal subgroup, so we can form the quotient group 
$G/G^\prime$, which turns out to be abelian and which we call the {\em
  abelianization} of $G$. Its importance lies in the fact that
$G/G^\prime$, equipped with the natural projection map $\pi': G \to G/G^\prime$,
is universal among
abelian groups $A$ equipped with a map $\sigma: G \to A$. In other
words, given any such $A$ and $\sigma$, there exists a unique map
$\sigma^\prime$ such that\footnote{Proof. $G/G^\prime$ is abelian since $ghG^\prime =
  ghh^{-1}g^{-1}hg G^\prime = hg G^\prime$. Given $\sigma$, define
  $\sigma^\prime (gG^\prime) := \sigma (g)$. This is well-defined since if
  $h G^\prime = g G^\prime$, then there exists a commutator $ j k j^{-1} k^{-1}$ (or a product of commutators) such that
  $h= g j k j^{-1} k^{-1}$. But then $\sigma^\prime (h G^\prime)
= \sigma (h) = \sigma (g j k j^{-1} k^{-1}) = \sigma (g) = \sigma^\prime
(gG^\prime)$ (a similar argument works for a product of generators). It is unique because if $\sigma^{\prime \prime}$ is another
such function, we must have $\sigma^{\prime \prime} (g G^\prime) =
\sigma^{\prime \prime} \pi (g) = \sigma (g) = \sigma^{\prime} \pi (g) = \sigma^{\prime} (g
G^\prime)$.} \footnote{In high-falutin' terms, we might say that
abelianization is a functor from the category of groups to the
category of abelian groups that is left adjoint to the functor that
forgets that a group is abelian.}
\begin{gather}\label{ump}\sigma = \sigma^\prime \circ \pi^\prime.
\end{gather}

To go further, we notice that the map $g \mapsto (\mathrm{\det}
\rho (g))^m$ defines not just a homomorphism, but also a character,
meaning that the heavy machinery of the theory of
characters can be brought to bear. Recall that, given a representation
$\rho$, the {\em character} $\chi$ afforded by $\rho$ is the map $G \to
\mathbb{C}: g \mapsto \mathrm{tr} \rho (g)$ defined by taking the
trace of the linear operator $\rho (g)$. Its {\em degree} is given by
its value on the identity element in $G$ or equivalently by the
dimension of the vector space carrying the representation
$\rho$. Characters of degree one, also called {\em linear
  characters}, are special because we then have, colloquially, that `$\mathrm{tr} \rho = \rho$';
as a result they land in $S^1 \subset \mathbb{C}$ and define
homomorphisms $G \to S^1$. 

More precisely, there is a one-to-one correspondence between
linear characters and homomorphisms $G \to S^1$ and this allows us to study
anomalies in complete generality. Indeed, the total number of linear
characters of $G$ is finite (being of degree one, they are necessarily
irreducible characters, so their number is bounded above by the number
of irreducible characters, which is equal to the number of conjugacy
classes, which in turn is bounded above by the number $|G|$  of
elements of $G$, which is finite). Moreover, as we shall soon see, the corresponding
maps $G \to S^1$ are easily explicitly determined, as are their kernels and
images. To study the anomalous properties of any particular
representation $\rho$, we merely need to explicitly match up
$\pi_{\rho}^m$ with a linear character on our list. This too
is easily done. In fact, because every representation of a finite
group is completely reducible, it is often possible to make statements
about the anomalies of all representations at once, as we shall see
when we discuss some examples. 

Before we describe the gory details, it is perhaps useful to
spell out exactly what it achieves in terms of allowing us to analyse the anomalies
of a specified representation $\rho$ of a specified group $G$. To do that, we need to ask what it
might actually mean to `specify' $\rho$ and $G$. 

At the most explicit level,
we might suppose that we are given (perhaps as a result of inspecting
a lagrangian of physical interest) a set of generators and relations for
$G$ along with a set of matrices, one for each generator, forming
a representation of $G$. Here one might think that our methods do not
bring much, since one can directly compute the $m$th powers of the determinants
of matrices and compare with unity. But even
here, our methods offer a slight advantage, in that we need only carry
out such a computation on a set of generators of the abelianization
$G/G^\prime$.\footnote{A set of generators of minimal size has one
  generator for each factor of
  $G/G^\prime$ expressed as $\Pi_i
\mathbb{Z}_{k_i}$, where each $k_i$ divides $k_{i+1}$.}

Less explicitly, we might be given the character
table for $G$ and the character $\chi$ afforded by $\rho$. Here, {\em a priori}, we would need to first reconstruct
$\rho$ and then compute its determinants (on
conjugacy classes) as above. With our methods, we can extract the
anomaly from $\chi$ directly. Of course, since the character table
gives us every irreducible character, we could also
figure out the anomaly of any representation (or its character) by reducing it. This
is made particularly straightforward by the fact that the determinant
of a reducible character is given by the product of the determinants
of its summands.

In fact, it is not even necessary to
know the character table, since we can (and will shortly show) reconstruct
the necessary part of it (namely the linear characters) directly. So
let us suppose we are in the worst-case scenario where we know $G$
only in the form of some generators and relations and are interested
in an arbitrary representation. Our first goal is to find the linear
characters. By the universal property in (\ref{ump}), these can be obtained by precomposing
the linear characters of the abelianization $G/G^\prime$ with the map
$\pi^\prime$. So we use the given generators and relations for $G$ to
find $G^\prime$ and then $G/G^\prime$, which we express as a product
of cyclic groups, i.e.
$G/G^\prime \cong \Pi_i
\mathbb{Z}_{k_i}$, where each $k_i$ divides $k_{i+1}$, and choose a
generator for each factor. The
$|G/G^\prime|$ distinct linear
characters of $G/G^\prime$ are then obtained by assigning a $k_i$-th
root of unity to the $i$th factor. 
Precomposing these with
$\pi^\prime$ then determines the linear characters on $G$.\footnote{We
  note that
the map (of sets) from the set of conjugacy classes of $G$ to its
abelianization defined by $[g] \to gG^\prime$ is surjective, but not
injective unless $G^\prime$ is the trivial group.}

From here, it is easy to answer the question of which anomaly free
subgroups arise when we consider the set of all possible representations. This
amounts to considering which linear characters arise in the image of
$\pi_{\_}^m$ and is easily settled since 
$\pi_{\_}^1$ obviously surjects onto the linear
characters. So the map $\pi_{\_,}^m$ hits precisely the linear
characters that are $m$-fold products of linear characters (with
multiplication of characters defined pointwise in the target). The
kernels of these yield the possible anomaly-free subgroups. 

For a second question that is easy to answer, suppose we are given
just one character $\chi$ and we wish to compute the anomaly free
subgroup of a representation afforded by it. 

Given $\chi$ we may define a
linear character as its
{\em determinant} $\mathrm{det} \chi : G \to S^1: g \mapsto  \mathrm{det} \rho (g)$, where $\rho$ is any representation
affording $\chi$. This may be computed explicitly from $\chi$ as
follows. Supposing $\chi$ has degree $n$, then we have the following beastly
formula relating the determinant of the character of an element to the characters of
the element's powers (obtained straightforwardly from the formula for an
arbitrary square matrix given in \cite{Kondratyuk:1992he}):
\begin{gather}
\mathrm{det} \chi (g) = \sum_{\{k_1, \dots k_l \in \mathbb{N}  | \sum l
  k_l = n\}}\Pi_{l=1}^n \frac{(-1)^{k_l+1}}{l^{k_l}k_l!} (\chi (g^l))^{k_l}.
\end{gather}
For $n=2$, for example, we have
\begin{gather}
\label{eq:2}
\mathrm{det} \chi (g) = \frac{\chi^2(g) - \chi (g^2)}{2}
\end{gather}
while for $n=3$ we have
\begin{gather}
\label{eq:3}
\mathrm{det} \chi (g)= \frac{\chi^3 (g) - 3\chi (g^2) \chi (g) +2\chi (g^3)}{6}.
\end{gather}
With the determinant in hand, we can simply raise it to the desired
power $m$ and read off the kernel to extract the anomaly-free
subgroup.

With power tools in hand, let us now apply them to find all the anomaly free
subgroups of irreducible characters for various groups that
have appeared in the literature on
flavour physics. 

We start with the group $S^3$ of permutations of three objects. It has 3 conjugacy
classes, labelled by the cycle lengths $(.),(..)$, and $(...)$. The
derived subgroup is isomorphic to 
$\mathbb{Z}/3\mathbb{Z}$, consisting of the union of the conjugacy
classes $(.)$ and $(...)$ and the map $\pi'$ sends those
classes to the trivial element $1 \in
\mathbb{Z}/2\mathbb{Z} \cong G/G^\prime$ and sends the class
$(...)$ to the non-trivial element $-1$. The 2 linear characters of
$\mathbb{Z}/2\mathbb{Z}$ are defined by $\chi_0 (1) = \chi_0(-1)=1$
and $\chi_1 (1) = 1, \chi_1(-1)=-1$. After precomposing with
$\pi^\prime$, we find the linear characters of $S_3$ given by $\chi_0
((.))=\chi_0 ((...)) = \chi_0((..))=1$
and $\chi_0
((.))=\chi_0 ((...)) = 1, \chi_0((..))=-1$. There is one further
irreducible character, which can be found using orthogonality to be
$\chi_2
((.))=2,\chi_2 ((...)) = -1, \chi_2((..))=0$.
Its determinant can only differ from 1 on the conjugacy class $(..)$, for which (\ref{eq:2}) yields $\mathrm{det} \chi_2 ((\cdot
    \cdot) ) = \frac{\chi_2^2 ((\cdot
    \cdot) ) - \chi_2 ((\cdot \cdot)^2)}{2} = \frac{0  - 2}{2} = -1.$
  Thus we identify
$\mathrm{ker} \det \chi_2 = \chi_1$. So for odd $m$, the
irreducible representations affording $\chi_1$ and $\chi_2$ are anomalous with
anomaly free subgroup $S_3^\prime \cong \mathbb{Z}/3\mathbb{Z}$, while
for even $m$ all representations are anomaly free.

This example is somewhat boring since the list of possible
anomaly-free subgroups has just 2 entries, {\em viz.} $S_3$ and
$S_3^\prime$. A more interesting example from flavour physics \cite{Frampton:1995fta} is the
quaternion group, of order 8. Henceforth we take 
the liberty of starting from the
character table, which can be found these days at
the touch of a button using {\tt GAP} \cite{GAP4}. The table for $Q_8$ is shown at the top
left in
Fig.~\ref{tab}.
\begin{figure*}
\begin{center}
    \begin{minipage}[b]{.49\linewidth}
      \begin{center}
        \begin{tabularx}{0.9\textwidth}{X|XXXXX|X}
$Q_8$ & 1 & -1& i& j & k &  $\mathrm{det}$ \\ \hline
$\chi_0$& 1 & 1 & 1 & 1 & 1 & $\chi_0$\\
$\chi_1$& 1&1&1&-1&-1&$\chi_1$\\
$\chi_2$& 1&1&-1&1&-1&$\chi_2$\\
$\chi_3$& 1&1&-1&-1&1&$\chi_3$\\
$\chi_4$& 2&-2&0&0&0&$\chi_0$
\end{tabularx}\\
      \end{center}
     \end{minipage}
   \begin{minipage}[b]{.49\linewidth}
     \begin{center}
      \begin{tabularx}{0.9\textwidth}{ X | X X X X| X }
$A_4$ & (.) & (12)(34)& (123)& (132) & $\mathrm{det}$ \\ \hline
$\chi_0$& 1 & 1 & 1 & 1 & $\chi_0$\\
$\chi_1$& 1 & 1 & $\omega$ & $\omega^2$ &$\chi_1$\\
$\chi_2$& 1 & 1 & $\omega^2$ & $\omega$ &$\chi_2$\\
$\chi_3$& 3 & -1 & 0 & 0 & $\chi_0$ \\
 &  & & &  & 
\end{tabularx}\\
     \end{center}
   \end{minipage}
\end{center}
 \begin{center}
\begin{tabular}{ c | c c c c c c c| c }
$SL(2,F_3)$ & $\begin{pmatrix} 1&0 \\0 &1 \end{pmatrix}$ &
                                                           $\begin{pmatrix} -1&0 \\0 &-1 \end{pmatrix}$& $\begin{pmatrix} 1&-1 \\0 &1 \end{pmatrix}$& $\begin{pmatrix} 1&1 \\0 &1 \end{pmatrix}$ & $\begin{pmatrix} 0&-1 \\1 &0 \end{pmatrix}$ &  $\begin{pmatrix} -1&1 \\0 &-1 \end{pmatrix}$& $\begin{pmatrix} -1&-1 \\0 &-1 \end{pmatrix}$&  $\mathrm{det}$ \\ \hline
$\chi_0$& 	1	&1&	1&	1&	1&	1&	1 & $\chi_0$\\
$\chi_1$& 	1&	1&	$\omega^2$&	$\omega$&	1&	$\omega^2$&	$\omega$& $\chi_1$\\
$\chi_2$&  1&	1&	$\omega$&	$\omega^2$&	1&	$\omega$&	$\omega^2$& $\chi_2$\\
$\chi_3$&  	2&	-2&	-1&	-1&	0&	1	&1& $\chi_0$\\
$\chi_4$&  	2&	-2&	$-\omega$&	$-\omega^2$&	0	&$\omega$&	$\omega^2$& $\chi_1$\\
$\chi_5$&  2&	-2&	$-\omega^2$&	$-\omega$	&0&	$\omega^2$	&$\omega$& $\chi_2$\\
$\chi_6$&  3&	3&	0&	0	&-1&	0&	0 & $\chi_0$
\end{tabular}
\end{center}
\caption{\label{tab} Character tables for $Q_8$ (top left) $A_4$ (top right), and
  $SL(2,F_3)$ (bottom) with conjugacy classes in the top row labelled by
  representative group elements. The characters are labelled in the
  left-hand column and their respective determinants appear in the right-hand column.}
\end{figure*}
A glance at the linear characters shows that the abelianization is
isomorphic to the Klein group $\mathbb{Z}/2\mathbb{Z}\times
\mathbb{Z}/2\mathbb{Z}$ of order 4, so the derived subgroup is of
order 2 and is the subset $\{1,-1\} \subset Q_8$.\footnote{We shall not need it for our
  trifles, but a useful fact in this context \cite{Isaacs} is that $g
\in G$ is a commutator (ergo a generator of $G^\prime$) iff. $\sum_i
\chi_i (g)/\chi_i (1) \neq 0$, where $i$ indexes the irreducible
characters.} But the three possible anomaly-free subgroups
for $m$ odd are larger (because the anomalous quotient must be cyclic), being all isomorphic to  $\mathbb{Z}/4$. One
is $\{\pm 1,\pm i\}$ and the other two are obtained by the outer
automorphisms of $Q_8$, which permute $i,j,$ and $k$. For $m$ even,
all representations are anomaly free.

For the group $A_4$, of order 12, of alternating
permutations of 4 objects, used in {\em e.g.} \cite{Ma:2001dn,Altarelli:2005yp}, the character table appears at the top
right in
Fig.~\ref{tab}.
The 3
linear characters indicate that $A_4/A_4^\prime \cong
\mathbb{Z}/3\mathbb{Z}$. Thus $|A_4^\prime|=4$ and
inspection of $\chi_1$ shows us that $A_4^\prime = [(.)] \cup [(12)(34)] \cong
\mathbb{Z}/2\mathbb{Z} \times \mathbb{Z}/2\mathbb{Z}$.
For the remaining
character (of degree 3), it suffices to compute the determinant on the
permutation $(123)$. Since $\chi_3 ((123)) = 0$, our formula collapses
to $\mathrm{det} \chi_3  ((123)) = \frac{2\chi_3   ((123)^3)}{6} =
\frac{\chi_3   ((.))}{3} = 1$, so $\mathrm{det} \chi_3 = \chi_0$. For
$m=1,2\ \mathrm{mod}\ 3$, the irreducible representations affording
$\chi_1$ and $\chi_2$ are anomalous with
anomaly-free subgroup $A_4^\prime \cong \mathbb{Z}/2 \mathbb{Z} \times
\mathbb{Z}/2 \mathbb{Z}$. 

For the group $SL(2,F_3)$, of order 24, consisting of $2\times 2$ matrices with elements in the
field $F_3$, used in \cite{Carr:2007qw,Feruglio:2007uu}, the character table appears at the bottom of Fig.~\ref{tab}.
We immediately read off that the abelianization is isomorphic to
$\mathbb{Z}/3\mathbb{Z}$. The derived subgroup thus has order 8 and
must consist of the union of the conjugacy classes containing $\begin{pmatrix} 1&0 \\0 &1 \end{pmatrix}$,
$\begin{pmatrix}
 -1&0 \\0
 &-1 \end{pmatrix}$
 and $\begin{pmatrix} 0&-1 \\1 &0 \end{pmatrix}$, which is isomorphic
 to $Q_8$. To compute the determinants, it is sufficient to consider the
conjugacy class containing $\begin{pmatrix} 1 & 1 \\ 0 &
  1\end{pmatrix}$. A simple calculation then yields the right hand
column in the table. For $m$ a multiple of 3, all representations are
anomaly-free; otherwise, only the irreducible representations with
characters $\chi_{0,3,6}$ are anomaly free, while the others have
anomaly free subgroup $SL(2,F_3)^\prime \cong Q_8$. Those who care for
such things may now continue {\em ad nauseam}.

Finally, we must return to the thorny question of whether there might
exist further anomalies, undetected by Fujikawa's argument. Such
anomalies, which necessarily can arise only if we consider spacetimes
with non-trivial topology, are signalled 
by a non-trivial value of the exponentiated
Atiyah-Patodi-Singer eta invariant. This is a bordism invariant for
finite $G$, but it is not known how to compute it in
 general (for cyclic groups, see \cite{Garcia-Etxebarria:2018ajm,Hsieh:2018ifc}). 

Even if it is non-trivial, it may be possible to cancel the resulting
anomalies without changing the degrees of freedom, by coupling to a
topological quantum field theory. Again, it is not known what form such
a theory may take, in general. One well-understood class of examples
amounts to replacing $G$ by an extension $\tilde{G}$ (and $\rho$ by its `restriction'
along $\tilde{G}\twoheadrightarrow G$), where the r\^{o}le of the topological theory is
to obstruct the lifting of an arbitrary principal $G$-bundle to a
principal $\tilde{G}$-bundle \cite{Garcia-Etxebarria:2018ajm,Hsieh:2018ifc}. Again, we may return to Iba\~{n}ez
and Ross \cite{Ibanez:1991hv} for a pertinent example, where there exists an $l \in
\mathbb{N}$ such that the extension
$\mathbb{Z}/nl\mathbb{Z} \twoheadrightarrow \mathbb{Z}/n\mathbb{Z}$ can be used to cancel the pure gauge anomaly
for $\mathbb{Z}/n\mathbb{Z}$ \cite{Banks:1991xj}. Our arguments show
that this cannot happen for the anomaly that we have considered: any such anomaly
corresponds to a linear character $G \to S^1$, which cannot be sent to
the trivial character by pullback along any extension $\tilde{G} \twoheadrightarrow G$.
A similar argument from the bordism point of view appears in \cite{Hsieh:2018ifc}. 

Acknowledgment: I am grateful to Joe Davighi for a discussion. This work was supported by STFC consolidated
grant ST/T000694/1.

\end{document}